# FASTER THAN LIGHT COMMUNICATION : IS IT POSSIBLE ?

Leonardo Chiatti
AUSL VT Medical Physics Laboratory
Via Enrico Fermi 15, 01100 Viterbo (Italy)

July 2008

**Abstract**

A quantum optical apparatus permitting a faster than light communication between distant locations has been recently proposed by Shiekh. Some severe conceptual difficulties concerning this proposal are briefly addressed.

**The Shiekh argument**

In a recently published paper Shiekh (1) proposes an experimental apparatus permitting, according to his opinion, a faster than light transfer of information (but not of energy or matter) between distant locations. This apparatus is illustrated in Fig. 1.

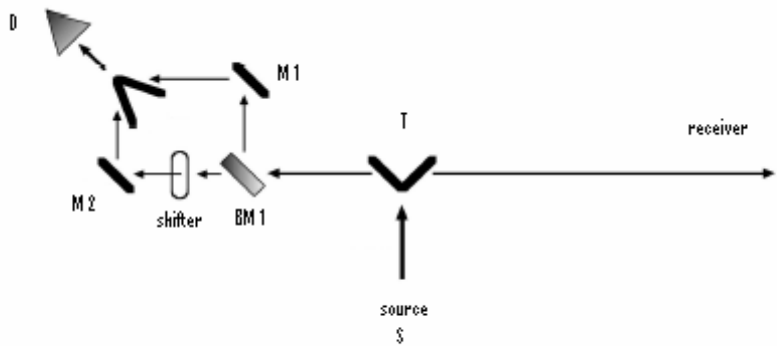

Fig.1 ; The Shiekh original apparatus.

Single photons emitted by a light source S interact with a mirror T, undergoing to a reflection towards a remote receiver or, alternatively, towards a Mach-Zender interferometer; the probabilities of these two results are assumed to be equal.
A phase shifter is placed across one of the interferometer arms, and the induced phase shift is assumed to be adjustable.
Let us suppose to adjust the shifter in such a way to obtain a destructive interference between the waves propagating along the interferometer arms $BM_1$-$M_1$-D and $BM_1$-$M_2$-D. Under these circumstances, the photodetector D should not detect any photon at all. Therefore, Shiekh assures, no photon actually travels the interferometer arms or, in other words, no photon is directed by T towards the interferometer. Thus, we have to do with an event (the shifter adjustment) able to modify the emission properties of the source S *instantaneously* !

Immediately after this event has occured, all the photons emitted by S are reflected by T towards the remote receiver, so that an enhancement of the incoming photonic flux is measured by the receiver. The information about the shifter adjustment is thus propagated from the shifter itself to the receiver, this propagation being faster than light.

**Some possible flaws**

If we accept the causality principle, it is impossible to modify the past as well as the "elsewhere" (that is, the spacetime zone external to the light-cone having its origin here and now). Of consequence, the shifter adjustment is unable to modify the characteristics of the photon emission by S : the absolute number of photons coming from S and directed towards the interferometer (or the remote receiver) remains unaffected by such an operation. The photonic flux at the receiver remains the same and no faster than light communication actually happens.

The real problem is : how explain the disappearing of the photons entering into the interferometer when the shifter is adjusted to obtain a destructive interference? In order to understand this problem, we must consider that normally a Mach-Zender interferometer operates with two detectors $D_1$, $D_2$ instead of just one (see Fig. 2).

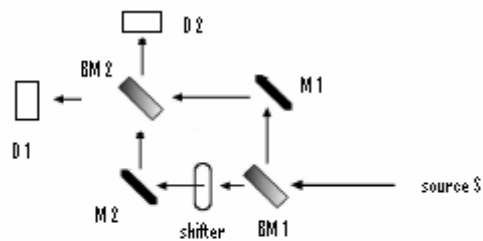

Fig. 2; Mach-Zender interferometer with two photodetectors.

The light rays reflected by mirrors $M_1$, $M_2$ do not incide upon a single detector D after a recombination; rather, they incide upon a second beamsplitter $BM_2$. In the zone between $BM_2$ and $D_1$ exists a superposition of the light beam reflected by $M_1$ and successively transmitted by $BM_2$ with the light beam reflected by $M_2$ and successively reflected by $BM_2$. Analogously, in the space between $BM_2$ and $D_2$ exists the superposition of the light beam reflected by $M_1$ and successively reflected by $BM_2$ with the light beam reflected by $M_2$ and successively transmitted by $BM_2$.

By adjusting the shifter is possible make one of these superpositions null but, in this case, the other superposition is maximized. Thus, we can make the output signal from detector $D_1$ ($D_2$) null, but then the output signal from $D_2$ ($D_1$) is maximum. A purely destructive interference never occurs, because the electromagnetic field is null only in the space between $D_1$ and $BM_2$ or, alternatively, in the space between $D_2$ and $BM_2$.

According to the experimental setup suggested by Shiekh, recombined waves incide directly upon a single detector D (Fig. 1). Therefore an interference pattern, with maxima and minima, exists in the zone where D is placed. If D size is sufficiently small, then it can be placed in a point where the elecromagnetic field is null, so that no photon will be detected; but this does not means the absolute absence of photons : they simply will not interact with D, continuing their travel to infinity.

Instead, if the detector D is sized big enough to cover the entire interference zone, with its maxima and minima, all the incident photons will be unavoidably detected and no condition of "destructive interference" will be possible.

Even classically, indeed, the field energy averaged on the entire interference zone exactly equates the sum of energies of each single component field, averaged on the same region. This well known result (energy conservation law) strictly assures no photon is lost due to the interference, no matter how the shifter is adjusted.

This counterargument is substantially equivalent to that of Bassi and Ghirardi (2), criticized by Shiekh in (3). Yet, a subtle but important difference is here the heuristic role played by the *local* character of energy conservation. In fact, there are only two possibilities to obtain a purely destructive interference as requested by Shiekh.

1) The energy conservation is locally violated. Thus, let us suppose to work with a strictly timed photon emission. By executing the requested shifter adjustment *after* the passage of a single photon in both the interferometer arms but *immediately before* the recombination occurs, this event should involve the photon disappearing. Alternatively, adjusting the shifter *after* the photon emission but *before* its passage in the interferometer, the photon disappears. We think that, unless the disappearing of single, free photons in flight is evidenced experimentally, this hypothesis must be rejected.

2) The energy conservation locally holds in the interference area, but the shifter adjustment retroacts on the photon emission event in the past, cancelling or modifying it. This seems the solution preferred by Shiekh.

Accepting this solution, however, the energy conservation is violated *inside the source*. In fact, a change in the emission characteristics of the source presupposes a change of the atomic or molecular forces acting inside it. This change should take place in order to maintain an external constraint (according to Shiekh, the total number of emitted photons is unaffected by the shifter adjustment, but their direction changes) arbitrarily fixed, so that a violation of conservation laws is generally unavoidable.

Finally, we remark that both the possibilities involve an alteration of the photon physical state induced by the shifter adjustment *after* the shifter-photon interaction has occured or *before* it can take place. This magical action-at-distance cannot be accepted.

**Conclusions**

Shiekh argument seems to be flawed due to an incorrect definition of the proposed experimental setup. It rest upon the achievement of a particular physical condition (the "destructive interference") by means of a suitable adjustment of the apparatus, a condition which is not actually feasible.
Therefore, no instantaneous action at distance on the light source really happens, and no faster than light communication arises.

**Bibliography**

1. A.Y. Shiekh; EJTP 5 No. 18 (2008) 105-108; arXiv:0710.1367

2. A. Bassi, G.C. Ghirardi; arXiv:0711.4538

3. A.Y. Shiekh; EJTP 5 No. 18 (2008) 109-112